\documentclass[12pt]{article}
\usepackage{color}
\usepackage{amssymb}
\usepackage{epsfig}
\usepackage{footnote}
\usepackage{longtable}
\usepackage{verbatim}
\usepackage{array,multirow}
\usepackage{titlesec}
\usepackage[fleqn]{amsmath} 
\usepackage{graphicx}
\usepackage{caption}
\usepackage{epstopdf}
\include{symbols}
\def\LAPPDTM{LAPPD$^{TM}$}
\def\p4D{precision-4D}
\def\P4D{Precision-4D}
\def\SBNF{Short Baseline Neutrino Facility~}
\def\FTBF{Fermilab Test Beam Facility~}
\begin{document}
\pagestyle{plain}
%


%
%

\begin{center}
{\Large\bf Measuring the Neutrino Event Time in Liquid Argon by a
  Post-Reconstruction One-parameter Fit}\\

{\it A White Paper for Snowmass 2021}\\
\today
\end{center}

\begin{center}
E. Angelico, A. Elagin, H. J. Frisch, \\
{\it Enrico Fermi Institute, the University of Chicago}\\
M. Wetstein\\
{\it Iowa State University}
\end{center}
%
%
\begin{abstract}
We propose a relatively simple method to measure the event time in liquid Argon (LAr)
TPC-based neutrino detectors that takes advantage of the topological reconstruction of
each event from the TPC data prior to performing a `one-parameter' fit. Measured times and
positions of detected photons are fit to the expected pattern of light from the tracks as reconstructed using the electron drift. The event can be treated as a rigid body with only the neutrino interaction time as a free parameter. The optical properties of LAr are comparable to those of water for Cherenkov light in visible wavelengths. Data-Monte Carlo comparisons of the light patterns, given the known track topology from electron drift, enable {\it in situ} calibration of the optical model and further optimization of the timing. A back-of-the-envelope calculation predicts that the single parameter fit for the interaction time requires a significantly lower photodetector coverage than needed for the same precision in conventional warm-liquid detectors.


\end{abstract}

\vskip-0.5in

%
%

\section{Introduction}

The proposal to rebunch the Fermilab Main Injector beam on a higher RF
harmonic~\cite{BeamTiming_PRD}  would allow both the LBNF Near and Far
detectors~\cite{LBNF} to see the full on-axis fluxes binned by neutrino arrival time. Each
time bin serves as a contemporaneous Far/Near-detector oscillation experiment with its own
characteristic flux spectrum selected on the neutrino time-of-arrival. However,
instrumenting the Far detector with fast timing has been regarded as daunting due to the
number of photo-detectors required for large fractional coverage. Here we describe a
method that exploits an inherent advantage of TPC-based LAr detectors over large
warm-liquid neutrino detectors to substantially reduce the required coverage for
extracting the event interaction time at the needed precision.


The organization of this White Paper, intended as input to Snowmass 2021, is as follows.
Section~\ref{stroboscopic_method} recapitulates the `Stroboscopic' proposal for exploiting
the correlation of true neutrino energy with the detected neutrino time-of-arrival by
rebunching the Main Injector beam on a high RF harmonic~\cite{BeamTiming_PRD}.
Section~\ref{Photodetectors} gives a brief introduction to the necessary `4D-precision
photo-detection', here defined as the simultaneous measurement of photon arrival with a
time resolution of tens of picoseconds and a space resolution of several millimeters. The generation of Cherenkov light from the the charged particle
tracks reconstructed by the LAr TPC is described in Section~\ref{post_recon_simulation}.
Section~\ref{1P_fit} describes the 1-parameter fit of the neutrino arrival time using the time/space coordinates of simulated photons relative to the coordinates of detected photons.
Determining the photodetector coverage needed to identify the neutrino time bin, including
the use of mirrors to extend cathode coverage, is briefly discussed in
Section~\ref{coverage}. Section~\ref{conclusions} presents conclusions.


\section{The Stroboscopic Method}
\label{stroboscopic_method}

The Fermilab-Chicago Timing Planning Meeting workshop~\cite{Chicago_meeting} was held in
March, 2018 to discuss what currently inaccessible physics one could do with
`precision-4D' measurements~\cite{precision_4D}. The discussion led to a proposal to
rebunch the Main Injector beam on the 10th harmonic of the current RF frequency, 531 MHz,
to produce shorter proton bunches~\cite{BeamTiming_PRD}. Although the 120 GeV protons and
the neutrinos are ultra-relativistic, the difference in the velocity of the parent hadrons
from $c=1$ leads to a correlation of the arrival time of neutrinos at both the Near and
Far detectors with energy. In November 2019, Fermilab, Chicago, and Iowa State held a
second meeting at Fermilab, the Workshop on Precision Time Structure in On-Axis Neutrino
Beams~\cite{Fermilab_workshop, Matt_WandC}, to vet details with accelerator and neutrino
detector experts.

\begin{figure}[ht]
\centering
\includegraphics[angle=0,width=3.0in]{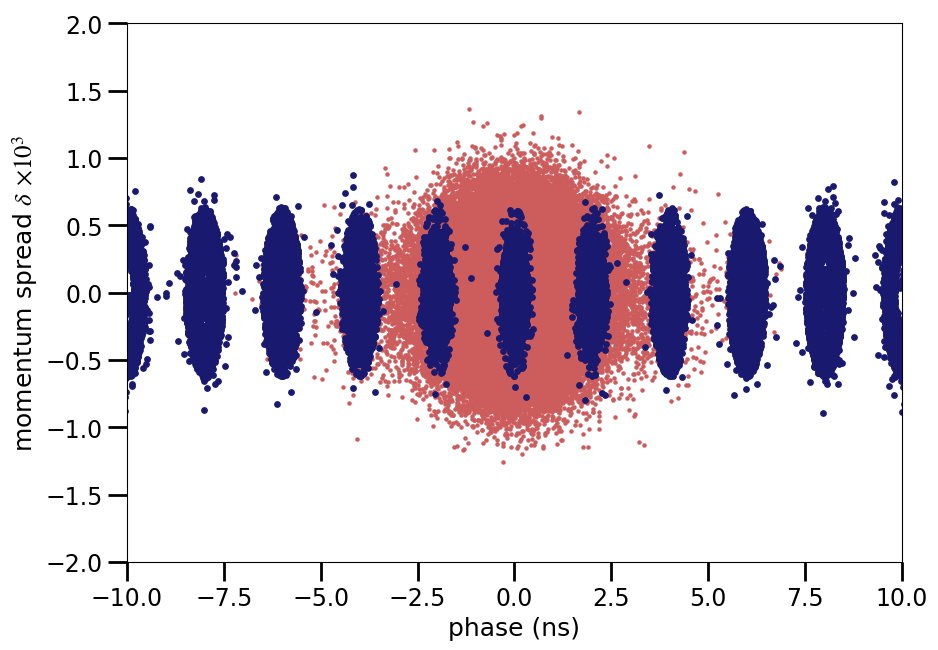}
\hfil
\includegraphics[angle=0,width=3.0in]{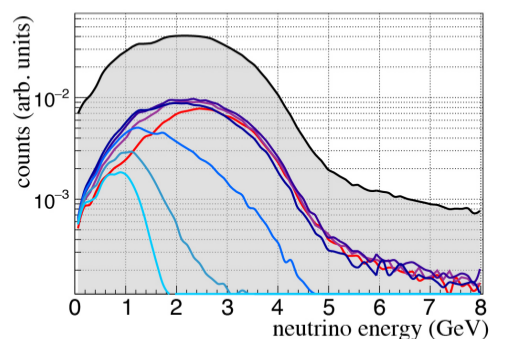}
\caption{Left: the momentum spread for one bunch of the current 53
MHz Main Injector beam versus the phase in nanoseconds (Red), and
for the beam rebunched on the 10th harmonic (Blue).
Right: the neutrino momentum spectra in 200 psec bins of time of
arrival relative to the proton bunch assuming a 100 psec detector
resolution and a 250 psec proton bunch width.}
\label{fig:stroboscopic_method}
\end{figure}

The left-hand panel of Figure~\ref{fig:stroboscopic_method} shows the
momentum spread in one bunch of the current 53 MHz Main Injector beam
versus the phase in nanoseconds in red, and the same for the beam
rebunched on the 10th harmonic in blue~\cite{BeamTiming_PRD}. The
right-hand panel shows the neutrino momentum spectra in 200 psec bins
of time of arrival relative to the proton bunch assuming a 100 psec
detector resolution and a 250 psec proton bunch width.  The
late-arrival bins have a much softer distribution in neutrino energy
than the early-arrival bins.

The beauty of this is in the correlation of true neutrino energy with an observable
unrelated to the neutrino energy as measured in the detector. Each time bin serves as a
contemporaneous experiment, viewing the identical detector, and undergoing the identical
identification of electron appearance, muon CC and NC events, tau neutrinos, and other
signatures, but with its own characteristic neutrino flux versus energy.

The Stroboscopic proposal~\cite{BeamTiming_PRD} allows both the Near and Far detectors to
see the full on-axis time-energy correlated fluxes contemporaneously. However,
instrumenting the Far detector with fast timing has been regarded as daunting due to the
number of phototubes required for large fractional coverage. Here we present a method that
exploits the advantages of LAr to substantially reduce the required coverage for
extracting the event interaction time at the needed precision.

\section{Large-Area High Space/Time-Resolution Photo-detection}
\label{Photodetectors}

The capability of simultaneously measuring the position of a charged particle or photon and the time of arrival at a precision comparable to light travel times of a millimeter, dubbed `Precision-4D', enables imaging of the event topology from photon travel times
given enough detected light. Systems of Silicon Photomultipliers (SiPMs) or MCP-based
photomultipliers (MCP-PMTs) such as the Incom LAPPD~\cite{Incom_production} would provide
adequate time and space resolution for exploiting the time-energy correlation. A possible
electronics architecture for the \p4D photodetector consists of a multi-buffer waveform
sampling front end~\cite{PSEC4A} with an effective buffer length long-enough to
accommodate the DUNE trigger latency~\cite{DUNE_trigger}.

Time at the Near and Far detector locations~\cite{DUNE_timing} and bunch-by-bunch at the
production target would be recorded relative to a master clock distributed via a system
such as White Rabbit~\cite{white_rabbit}. Synchronization among these three locations at
the required level is within the capabilities of current
technology~\cite{GPS_time_distribution}. The requirements for clock distribution within to
the photo-detectors within each of the Near and Far detectors are within the capabilities
of the current system at the Fermilab TestBeam Facility.

We note that the determination of event time is done after the event has triggered and has
been reconstructed, requiring no changes to the current plans for the optical detector
system for scintillation light or triggering.

\section{Optical Properties of Cherenkov Light in Liquid Argon}

Precision timing in liquid Argon necessitates the precision detection of prompt light from the neutrino interaction, and Cherenkov light in visible wavelengths has the properties needed to achieve 50-100 ps timing. The optical properties of Liquid Argon have been studied in visible wavelengths and its properties appear to be suitable for Cherenkov detection~\cite{{grace},{sinnock},{seidel},{bideau-mehu}}. The index of refraction is similar to that of water, roughly 1.22 at 500~nm, and with less chromatic variability. Scattering lengths in the visible spectrum exceed 100~m for wavelengths above 300~nm. Cherenkov detection in LAr-TPCs was successfully demonstrated by the ICARUS collaboration~\cite{ICARUScherenkov}, but has remained largely unexplored since.


\section{Cherenkov Pattern-of-Light Fitting with a Fully Reconstructed TPC Event Topology}

\label{post_recon_simulation}

 The reconstruction of the event time in large warm-liquid detectors such
as JUNO~\cite{Juno} and Hyper-Kamiokande~\cite{HyperK} requires a photodetector coverage
approaching unity for reconstructing the topology and locating the event vertex from
Cherenkov light, as well as for good energy
resolution~\cite{Andrey_paper_1,Andrey_paper_2,Andrey_paper_3}.  For a cryogenic detector the size of the DUNE Far detector, the photodetector cost and scope of the detector upgrades would be unmanageable. Chromatic dispersion and scattering complicate the picture; both hardware and software schemes can mitigate the effects, but there is a net information loss leading to a time resolution comparable or larger than needed for the stroboscopic timing in the Far detector. For many reasons instrumenting the Far detector in DUNE with Precision 4D for event reconstruction has been considered a non-starter~\cite{Fermilab_workshop}.

 However, the need for photodetector coverage in a LAr detector is different from that in conventional large
warm-liquid detectors in that the TPC provides a full precise topological reconstruction
of the event. Here we propose adding a dedicated sparse system of photo-detectors with
cm-scale time and space resolution~\cite{natural_units} to record photons from Cherenkov
light from the charged tracks in the event.  The output from the LAr event reconstruction
of the TPC data is used as input to a pattern-of-light simulation that generates Cherenkov light from
reconstructed tracks. The simulated photons are propagated to the detector inner surfaces
where position and associated time-of-arrival (hits) are recorded. The predicted 4D
`hits'-- the time of photon arrival and associated position-- are then compared to the
measured time-of-arrival and position of the measured hits. A measure of goodness-of-fit
plotted versus time yields a best-fit to the neutrino interaction time and its
uncertainty.

\subsection{The Event as a Rigid Body}
\label{rigid_body}

The precision of the reconstructed topology of a neutrino event in a very large LAr
detector is expected to be a few mm~\cite{Icarus_distortion}, i.e. on the order of 10 psec
light travel time~\cite{natural_units}. This is small compared to the Stroboscopic time
binning~\cite{BeamTiming_PRD}. The reconstructed event thus can be considered a `rigid
body' in the classical sense, i.e. having no internal degrees of freedom (DOF) on the
scale needed for event time reconstruction. At the required precision the 6 DOFs of event
position and orientation also should be adequately measured~\cite{Icarus_distortion} after
corrections.


\subsection{Simultaneous Optimization of Event Timing and Optical Model}
\label{photon_simulation}

The tracks in the reconstructed event from TCP drift provide a detailed input to a simulation to generate Cherenkov photons. The photons are then propagated to the detector surfaces and position and time relative to the primary neutrino vertex are recorded. Additional truth information such as photon wave-length, polarization, and scattering history is also known in the simulation. This enables calibration of the optical properties of the detector to further optimize the timing fits.


\section{The Post-Reconstruction One-parameter Fit}
\label{1P_fit}

\subsection{Fitting the observed 4D hit list to the simulated LAr hit list} \label{fitting}

 The position and time for Cherenkov photons generated by the simulation
from the reconstructed tracks will depend on the time of the event. For efficiency not all
Cherenkov photons need to be simulated; the generation may be limited to tracks only above
some momentum, length, or angle. The uncertainties in time and position of the simulated
photons may be parametric, depending on photon drift length, wavelength, direct or
reflected, and possibly event type and topology.

The two lists, measured and simulated, are then compared in
a 1-parameter fit, for example a simple ${\chi}^2$ fit versus
event time. Not all generated Cherenkov photons will have a match,
nor will all measured photons have a match within the estimated
error. Both the number of matches and the goodness-of-fit
enter into the determination of the event time.

\subsection{Chromatic dispersion, Absorption, and Polarization}
\label{chromatic_dispersion}

 The list of predicted hits and tracks used for fitting the
measured data can be curated; not all emitted Cherenkov photons need to used in the
comparison. For example, an initial fit could use only simulated photons in a limited
wavelength range to minimize dispersion. After the fit, other, un-matched hits, can be fit
as a function of wavelength to measure chromaticity. Similar information on photon
scattering, absorption, and polarization, for example, can be extracted from the data by
selecting on photon optical paths.


\section{Photodetector Coverage}
\label{coverage}

 The desired precision on the event time determines the photodetector
coverage. Both the distribution of photodetector modules and the overall fractional
coverage need to be optimized. A given `tiling' of the detector surfaces with
photo-detectors yields the number of matched hits versus event time. The match rate versus
coverage informs the scope of the required coverage.

\subsection{The Use of Mirrors and the Optical Time Projection Chamber}
\label{mirrors_OTPC}

\begin{figure}[ht]
\centering
\includegraphics[angle=0,width=0.25\textwidth]{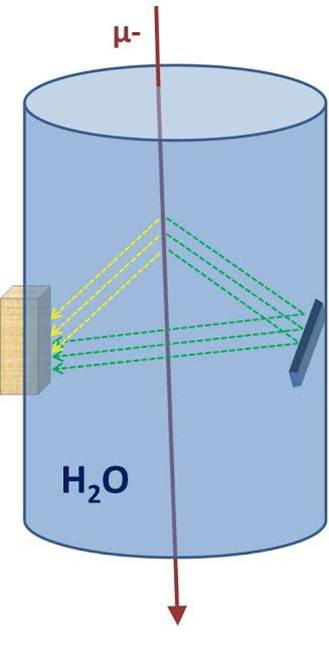}
\hfil
\includegraphics[angle=0,width=0.65\textwidth]{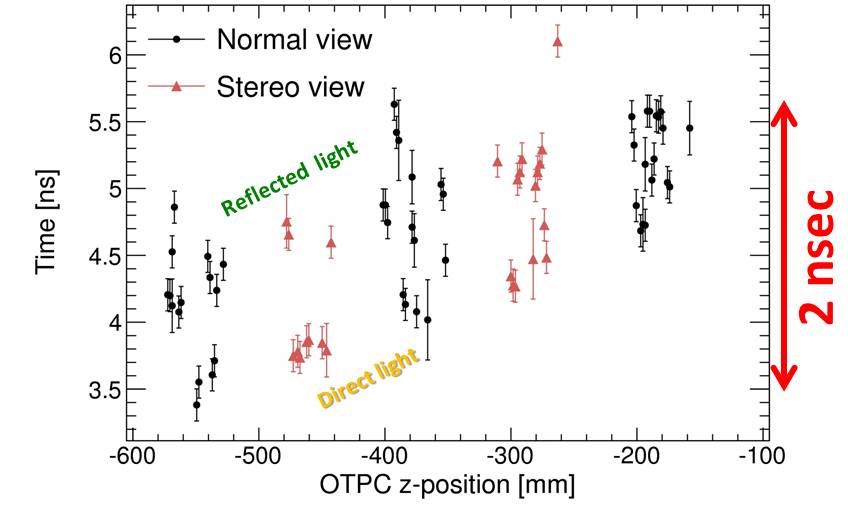}
\caption{Left:The Optical TPC Concept. The actual OTPC employed 5
  Planacons each with 30 measurements, arranged in a stereo
  configurations~\cite{OTPC_paper,Eric_thesis}. Right: Data from the
  traversal of a single muon through the OTPC, showing the time of
  arrival versus distance along the OTPC of hits in the Planacon
  MCP-PMTs.  The reconstructed `track' from direct light is earlier
  than that from reflected light by 785 psec.}
\label{fig:OTPC}
\end{figure}

\P4D detectors allow mirrors to multiply photocathode coverage by using the drift time of
photons to constrain the path-length travelled from the source for both direct and
reflected photons. Figure~\ref{fig:OTPC} describes the use of mirrors in a small Optical
TPC, which was tested in the Fermilab MCenter test beam~\cite{OTPC_paper,Eric_thesis}. The
 left-hand panel shows the concept. A plastic tube filled with water
 provides Cherenkov radiation for particles travelling down the tube;
 five Planacon~\cite{Planacon} MCPs detect the light in a 30-degree
 stereo configuration, with each MCP providing 30 points along the
 track via a micro-strip anode. Planar mirrors on the opposite side of
 the tube reflect light from the other side of the Cherenkov cone onto
 the MCP's, with the light arriving $\approx$ 785 psec later. The
 right-hand panel shows the time-of-arrival versus distance along the
 OTPC axis of hits in the Planacon MCP-PMTs for a single muon
 event. Two `tracks' are visible: the track from Cherenkov direct
 light is earlier than the track made by the light reflected by
 mirrors on the other side of the tube by 785 psec. The slope is
 consistent with the muon being fully relativistic. Even though this
 OTPC proof-of-principle prototype was very short, in a length of 40
 cm the measured angular resolution was 16
 mrad~\cite{OTPC_paper,Eric_thesis}.

\subsection{Far Detector Optical Considerations}
\label{geometry}

Substantial work has been done on the optical systems for DUNE~\cite{Himmel_1, Himmel_2}
and would not be affected by this proposal. Also one cannot be specific about a
low-coverage optical system or a new detector for Cavern 4 so far in the future. Detailed
simulations are essential to determine the dependence on time resolution to the number of
photons detected. However some general observations follow.

A low-coverage \P4D optical system relies on the detection of Cherenkov light in the
optical range by detector modules with time resolutions less than 100 psec. The detectors
need to operate at LAr temperature inside the cryostat to avoid penetrations~\cite{DaveS}.
One attractive possibility is SiPMs, possibly in strips in the gaps between the field
cages electrodes~\cite{DaveS}. Alternatively, it may be possible to leverage part of the existing ARAPUCA photodetection system, left bare without wavelength-shifter, for Cherenkov light detection in visible wavelengths. This would require no changes to the structure or feedthroughs of the existing design of the first LAr-TPC module.


 Mirrored
surfaces can be used to amplify cathode coverage and provide stereo information as in the
OTPC, including reflections from internal electrical structures~\cite{Philadelphia}.



\subsection{Machine Learning and More Sophisticated Fitting}
\label{machine_learning}

Optimizing the photodetector/mirror system coverage and distribution for different event
types is a many-parameter problem. It may be a suitable problem for machine learning.

\section{Conclusions}
\label{conclusions}


 The Stroboscopic method provides an
opportunity for the simultaneous measurement of neutrino oscillations using both Far and
Near detectors for each of the time-selected spectra in neutrino energy. Liquid
Argon-based detectors have an advantage over traditional warm-liquid detectors in that
each event is precisely reconstructed in space from the electron TPC data. The
reconstructed tracks then can be used to simulate the detected time and position of
Cherenkov photons radiated from some or all of the charged particles in the event. The
comparison of the 4D-coordinates of simulated photons to those of measured photons depends
on one parameter-- the time of the event, which can be fitted for.

Chromatic dispersion and other effects can be
addressed by limiting the matching to simulated photons in a selected
range of wavelengths from the `truth' information.

The power of the fit will depend on the number of matched hits, which in turn depends on
the photodetector coverage. However because one is not reconstructing the event, but
instead comparing predicted to measured 4D-hits, a much smaller required coverage is
expected. If so, the one-parameter fit method  would allow application of the Stroboscopic
method to the large Far detectors as well as the Near detector.

\section{Acknowledgements} We thank Dave Schmitz for
discussion of LAr geometries and constraints.



\begin{thebibliography}{99}
\bibitem{natural_units} We use the convention that $c=1$ so that
  time and space can be measured in meters.
\end{thebibliography}
\end{document}